\documentclass[aps,prl,twocolumn,groupedaddress]{revtex4}
\usepackage{graphicx}

\newcommand{\br}{\mathbf{r}}

\begin{document}

\title{Is there a Glass Transition in Planar Vortex Systems?}

\author{Thorsten Emig} 
\affiliation{Institut f\"ur Theoretische Physik,
  Universit\"at zu K\"oln, Z\"ulpicherstr. 77, 50937 K\"oln, Germany}

\author{Simon Bogner}
\affiliation{Institut f\"ur Theoretische Physik, Universit\"at zu
  K\"oln, Z\"ulpicherstr. 77, 50937 K\"oln, Germany}

\date{\today}

\begin{abstract}
  The criteria for the existence of a glass transition in a planar
  vortex array with quenched disorder are studied.  Applying a replica
  Bethe ansatz, we obtain for self-avoiding vortices the exact
  quenched average free energy and effective stiffness which is found
  to be in excellent agreement with recent numerical results for the
  related random bond dimer model \cite{Zeng+99}. Including a
  repulsive vortex interaction and a finite vortex persistence length
  $\xi$, we find that for $\xi \to 0$ the system is at {\em all}
  temperatures in a glassy phase; a glass transition exists only for
  finite $\xi$. Our results indicate that planar vortex arrays in
  superconducting films are glassy at presumably all temperatures.
\end{abstract}

\pacs{}

\maketitle 

The existence of a vortex glass (VG) in type-II superconductors is of
crucial importance for their performance.  Quenched disorder affects
both the translational order of the Abrikosov vortex lattice and the
low current response \cite{Nattermann+00}. The thermodynamic response
of a planar VG in a mesoscopic superconducting film \cite{Abrikosov64}
has been studied in a recent experiment \cite{Bolle+99}, providing
`fingerprints' of the sample's disorder \cite{Emig+00}.  Glassy phases
are expected to be generic to a wide class of disorder dominated
systems.  The planar VG is a famous starting point within this class
since it is among the very few systems where analytical progress seems
possible.  The situation is here analogous to that of exactly solvable
1D quantum systems which have advanced the understanding of strongly
correlated systems.

Thermal fluctuations compete against the VG, raising the question of
the existence of a glass transition at a finite temperature $T_g$
\cite{Fisher89}.  However, the literature on this subject is rather
contradictory \cite{Nattermann+91,Hwa+94,Zeng+99}.  Major support,
both analytical and numerical, for a finite $T_g$ stems from the
phenomenologically related random field (RF) XY model which has a
finite $T_g$ transition \cite{Cardy+82,Giamarchi+94}. However,
considering the entropic vortex interaction, the absence of a finite
$T_g$ transition has been predicted under certain assumptions as well
\cite{Nattermann+91}. The theory underlying the vortex system can be
applied to a broad class of other systems like surface steps
\cite{Mullins+64}, polymers \cite{DeGennes68} and domain walls
\cite{Pokrovsky+79}.  A more fundamental aspect of the planar VG is
related to its often debated replica description, including the
concept of replica symmetry breaking, and the non-trivial exchange of
replica and thermodynamic limit.  Due to the relation to the recently
simulated random bond dimer model \cite{Zeng+99}, the VG provides a
unique system to test replica theory and could be useful in clarifying
important concepts in the statistical mechanics of disordered systems.

In this paper, we analyze the conditions under which a planar vortex
lattice undergoes a finite $T_g$ glass transition. For vanishing
vortex persistence length $\xi$, we find the lattice to be glassy at
{\em all} temperatures with an analytic free energy. Analyticity of
the free energy is also obtained for the vortex-free RFXY model.  Our
replica symmetric Bethe ansatz results are found to be in excellent
agreement with recent simulations \cite{Zeng+99}. For finite $\xi$,
there exists a finite $T_g$ which, however, is presumably too large to
be observable in superconducting films.

{\em Vortex system.} --- We consider a lattice of self-avoiding
directed elastic lines of density $\rho=1/a$ in $1+1$ dimensions. Each
line is characterized by its position $x_j(z)$ and line tension $g$.
The lines interact via a short-ranged repulsive pair potential $U(x)$.
Quenched disorder couples locally to the lines via the random
potential $V(\br)$, $\br=(x,z)$, which is assumed to have short-ranged
correlations $\overline{V(\br)V(\br')}=\Delta \delta(\br-\br')$. The
total energy reads
\begin{equation}
\label{H-VS}
H=\int dz \sum_j \left\{ \frac{g}{2}
\left(\frac{dx_j}{dz}\right)^2+ \sum_{i,i\neq j} U(x_i-x_j)+
V(x_j,z) \right\}.
\end{equation}
This model can be considered as the continuum version of lines placed
on the bonds of a rectangular lattice \cite{Kardar+85}. Below, we will
also consider lines with a finite persistence length $\xi$,
corresponding to {\em correlated} random walks in the lattice model.

{\em Random field XY model.} --- The line lattice is usually treated
within a 2D elastic theory for the displacement field $u(\br)$ so that
the line positions are $x_j(z)=ja+u(ja,z)$. The effective Hamiltonian
is then
\begin{equation}
\label{H-XY}
H_{\rm XY}=\int\! d^2\!r \left\{\frac{c_{11}}{2}(\partial_x u)^2 +
\frac{c_{44}}{2}(\partial_z u)^2
+\rho(\br)V(\br)\right\}
\end{equation}
with compression modulus $c_{11}=a U''(a)$, tilt modulus $c_{44}=g/a$
and line density $\rho(\br)=\sum_j \delta(x-x_j(z))$.  Rewriting
$\rho(\br)$ as a Fourier series in $u(\br)$, the model becomes
equivalent to the random field XY model without vortices
\cite{Fisher89}. It is well known that, for weak disorder, this model
shows a phase transition to a glassy phase at a finite temperature
$T=T_g$. A renormalization group (RG) analysis yields the
scaling dimension of disorder \cite{Cardy+82},
\begin{equation}
\label{dim}
\lambda_{\Delta}=2\left(1-\frac{\pi T}{a^2 \sqrt{c_{11}c_{44}}}\right),
\end{equation}
giving $T_g=a^2\sqrt{c_{11}c_{44}}/\pi$. An important property of
the XY model is that the elastic moduli remain unrenormalized due to a
statistical symmetry \cite{Cardy+82,Hwa+94}. Thus, naively one would
expect for the line lattice a transition at a $T_g$ determined by the
elastic moduli on the length scale where $H_{XY}$ becomes applicable.
However, as we will show below, there is, in fact, a renormalization
of $c_{11}$ in the original model of Eq.~(\ref{H-VS}) which can even
destroy the transition.

What are the manifestations of the glass transition in the XY model?
Anomalous variations in the response to an external tilt of the lines
have been used to characterize the glassy phase \cite{Hwa+94}.
However, signatures of the transition in the free energy itself have
not been studied so far. Using the known mapping of the XY model onto
a vector Coulomb gas \cite{Cardy+82} one can derive the RG flow of the
free energy as shown in \cite{Kosterlitz74} for the pure XY model.
From this analysis we find that the quenched average free energy is
{\em analytic} at $T_g$.  Weak disorder makes the
contribution
\begin{equation}
f_\Delta=-\frac{\Delta}{T a^2 }\left(1+\frac{\Delta }{c_{11}^{3/2}
    c_{44}^{1/2} a^4}\frac{\pi^3}{(1-\tau)^2(1-2\tau)}\right)
\end{equation}
to the free energy density for small $\tau=1-T/T_g$, where we have
neglected $T$-independent terms.

Another signature of the glassy phase are its correlations
\cite{Nattermann+00}.  For $T>T_g$, one finds $[\langle (u(\br)-u({\bf
  0}))^2 \rangle ] \sim \log(|\br|/a)$, and in the glassy phase for
small $\tau>0$ the slightly faster growth $[\langle (u(\br)-u({\bf
  0}))^2 \rangle ] \sim \log^2(|\br|/a)$. Here $\langle\ldots\rangle$
and $[\ldots]$ denote thermal and disorder average, respectively.  The
decay of the corresponding density-density correlations suggests that
the correlation length is infinite on {\em both} sides of the
transition which is consistent with the predicted analyticity of free
energy. Next we will focus on the correlations of thermal fluctuations
$\delta u(\br)=u(\br)-\langle u(\br) \rangle$ around the state pinned
by disorder. On both sides of the transition one has
\begin{equation}
\label{corr-fct}
C(\br)=[\langle (\delta u(\br)-\delta u({\bf 0}))^2 \rangle ] = 
\frac{a^2}{16\pi K} \log(|\br|/a),
\end{equation}
with $K=(\pi/16)T_g/T=(a^2/16)\sqrt{c_{11}c_{44}}/T$. Below, we will
calculate $K$ directly in terms of the parameters of the original
model of Eq.~(\ref{H-VS}).

{\em Self-avoiding vortices.} --- A generic but simpler model is
obtained if the only vortex interaction of Eq.~(\ref{H-VS}) is the one
representing the non-crossing condition, i.e., a hardcore repulsion
$U(x)=c\delta(x)$ with $c\to \infty$. Using the replica method, with
$n$ replicas, the model can be mapped to $SU(n)$ fermions and thus can
be solved {\em exactly} by a replica Bethe ansatz (RBA) {\em without}
replica-symmetry breaking \cite{Kardar+85}. To date, only deep inside
the glassy phase, e.g., for $H \to H_{c1}$, the quenched average free
energy density $[f]$ has been studied by this technique
\cite{Kardar+85,Emig+01}.  However, at arbitrary $H$ (or density
$\rho$) the RBA yields
\begin{equation}
  \label{eq:fe-exact}
  [f]=f_0 \rho+
\frac{T^2}{2g}\int_{-Q}^Q q^2 \varrho(q) dq \, ,
\end{equation}
where $f_0$ is the single vortex free energy per unit length; its
precise form is unimportant in what follows.  
$\varrho(q)$ is
determined by
\begin{equation}
  \label{eq:int-eq}
  \int_{-Q}^Q \left\{ \frac{1}{l_d(q-q')} + 
    \pi \coth\left(\pi l_d(q-q')\right) \right\}\varrho(q') dq'=q
\end{equation}
with length scale $l_d=T^3/g\Delta$, and $Q$ is fixed by
$\int_{-Q}^Q \varrho(q) dq = \rho$.  The integral equation
(\ref{eq:int-eq}) can be solved perturbatively in $Q l_d$,
leading to
\begin{equation}
  \label{eq:rho-sol}
  \varrho(q)=\sqrt{1-(q/Q)^2}\left[ \frac{1}{2\pi} Q l_d -
\frac{\pi}{24} (Q l_d)^3 + \ldots \right].
\end{equation}
Higher order terms can be easily calculated and will depend also on
$q$, thus changing the functional form of the lowest order result. In
the opposite limit, $Q l_d \to \infty$, i.e, for vanishing
disorder, the solution is $\varrho(q)=1/(2\pi)$. With the result for
$\varrho(q)$ at hand, both $\rho l_d$ and the integral of Eq.
(\ref{eq:fe-exact}) can be obtained as a series in $(Q l_d)^2$.
However, we are actually interested in eliminating $Q$ in order to
calculate $[f]$ in terms of $\rho$ and $l_d$. At this stage, a
particularly interesting property of $\varrho(q)$ is important. The
{\em exact} result for the integral of Eq.  (\ref{eq:fe-exact}) as a
function of $\rho l_d$ is obtained from the {\em two lowest order
  terms} of Eq. (\ref{eq:rho-sol}). Indeed, it can be shown order by
order in $Q l_d$ that the exact result is
$\int_{-Q}^Q q^2 \varrho(q) dk = \frac{\pi^2}{3} \rho^3 + \rho^2
l_d^{-1}$.
This result we could also confirm numerically with high precision.
Thus, the RBA gives for the quenched average free energy density the
exact result
\begin{equation}
  \label{eq:f-final}
  [f]= f_0\rho+
\frac{\pi^2}{6} \frac{T^2}{g} \rho^3 + \frac{\Delta}{2T} \rho^2.
\end{equation}
For the interaction (last two) terms it is possible to give a simple
explanation. The first is just the known Pokrovski-Talapov interaction
\cite{Pokrovsky+79} from entropy loss due to collisions of thermally
fluctuating lines in pure systems. The second term can be explained by
a similar argument which gives the interaction in the dilute limit
$\rho \ll l_d^{-1}$.  It arises from an increase in energy due to
collisions which prevent the lines from gaining their optimal energy
from the random potential \cite{Kardar+85}. It is rather unexpected
that the total effective interaction is just the sum of these two
contributions. As a consequence of that, the free energy is an
analytic function. Based on our result for the XY model, this does not
exclude the possibility of a finite-$T$ glass transition. However, its
existence can be {\em excluded} by computing the scaling dimension
$\lambda_\Delta$, Eq.~(\ref{dim}), in terms of the macroscopic
compression modulus $c_{11}^0$. It is given by the effective
interaction, $c_{11}^0=aU_{\rm eff}''(a)$, with $U_{\rm eff}/a$
corresponding to the two last terms of Eq.~(\ref{eq:f-final}).  Using
$c_{44}=g/a$, and
\begin{equation}
  \label{eq:c-11-eff}
  c_{11}^0=\frac{\pi^2T^2}{ga^3} + \frac{\Delta}{Ta^2},
\end{equation}
we find $\lambda_\Delta>0$ at all finite temperatures, i.e.,
$T_g=\infty$. With the exact $c_{11}^0$ available, we are able to
relate the phenomenological stiffness $K$ of Eq.~(\ref{corr-fct}) to
the microscopic parameters of the model of Eq.~(\ref{H-VS}),
leading to
\begin{equation}
  \label{eq:K-eff}
  K=\frac{\pi}{16}\left(1+\frac{ag\Delta}{\pi^2T^3}\right)^{1/2}.
\end{equation}
For vanishing disorder, $\Delta=0$, the previously known result
$K=\pi/16$ is recovered \cite{Henley97}. 

There are a few comments in order to clarify the apparent discrepancy
between the exact RBA results for self-avoiding lines and the RG
predictions for the random field XY model of Eq.~(\ref{H-XY}).  To
begin with, on microscopic length scales there is no direct link
between the two models since $c_{11}=0$ due to self-avoidance as only
interaction. However, the XY model is generally believed to describe
the vortex lattice at length scales larger than $a$ where a finite
fluctuation induced $c_{11}$ exists \cite{Fisher89,Nattermann+91}.
That this point of view is incorrect for finding out about the
existence of a glass transition in the vortex system shows the
following argument. On large but finite length scales $L \gg a^2g/T$
the effective $c_{11}$ for $\Delta \to 0$ can be obtained from the
analogy of the self-avoiding vortex lines to 1D fermions
\cite{Kardar+85}. Since a finite vortex length $L$ translates into a
finite temperature for the fermions, the Sommerfeld expansion can be
used to get the correction $\delta c_{11}= -ag/3L^2$ to the result of
Eq.~(\ref{eq:c-11-eff}). Thus, on any sufficiently large length scale,
disorder seems to be {\em irrelevant} since then $\lambda_\Delta<0$.
In contrast, the RBA shows that disorder is always {\em relevant}. The
contradicting results are explained by the fact that the statistical
symmetry in the XY model is broken on {\em all} length scales by the
non-crossing condition.

\begin{figure}[th]
\begin{center}
\includegraphics[height=2.6in]{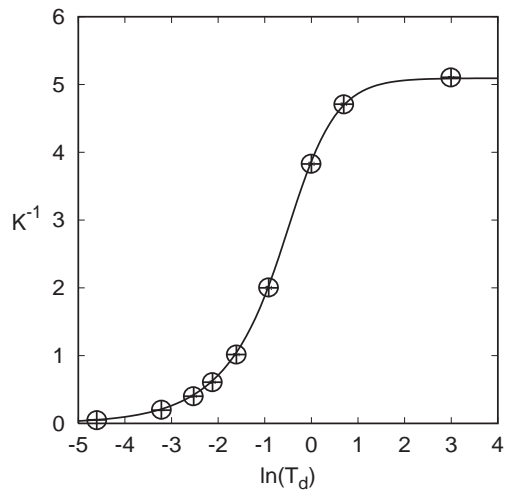}
\vspace*{-0.3cm}
\caption{Replica Bethe ansatz result for the stiffness $K$, 
  Eq.~(\ref{eq:K-eff}), plotted as function of the dimer temperature
  $T_d$, using Eq.~(\ref{T_d_of_Delta}) with $\xi_d=1.15$, and the
  corresponding data (circles) from the dimer model algorithm of Ref.
  \cite{Zeng+99}. 
}
\label{fig:1}
\end{center}
\end{figure}

{\em Dimer model.} --- The self-avoiding vortices can be simulated
directly, i.e., not via the XY model, using a mapping to a discrete
dimer model with quenched bond disorder \cite{Zeng+96}.  A recently
developed polynomial algorithm allows to study large systems at finite
temperature but fixed density $\rho=1/2$ (in units of the dimer
lattice constant) \cite{Zeng+99}. Comparison of our RBA predictions
with the results of this numerical approach provides a critical test
of the replica approach, including replica-symmetry and proper
analytical continuation while taking the number of replicas to zero.
The dimer model is defined as follows: Choose a subset (whose elements
are called dimers) of the bonds on a square lattice, such that every
lattice site is touched by exactly one of these dimers.  The energy of
one such dimer covering $D$ is defined by
$H_{d}=\sum_{(i,j)\in D} \epsilon_{ij}/T_d$,
where the sum is over all dimers of $D$, $\epsilon_{ij}$ are random
bond energies drawn from a Gaussian distribution with zero mean and a
variance of one, and $T_d$ is the dimer temperature. Via the
well-known mapping to the solid-on-solid model, it is possible to
assign every dimer covering a height function which maps to the line
displacement $u(\br)$ in the vortex system representation. This
representation results from the superposition of an arbitrary covering
$D$ and a fixed reference covering that specifies the direction of the
lines, cf.~Ref.~\cite{Zeng+99}.  By measuring the height
configurations, Zeng et al. could determine the correlation function
of Eq.~(\ref{corr-fct}), and thus the stiffness $K$ as a function of
$T_d$ \cite{Zeng+99}. In order to compare their results to our
prediction of Eq.~(\ref{eq:K-eff}), the parameters $g/T$ and
$\Delta/T^2$ have to be determined for the dimer representation of the
vortex system. It is easy to see that the ratio of horizontal and
vertical line steps on the lattice yields $g/T=2$.  The precise
relation between $\Delta$ and $T_d$ can be obtained by comparing the
{\em annealed} average free energy of the continuum model of
Eq.~(\ref{H-VS}) and the discrete dimer version of the vortex lattice.
Using known methods for computing the partition function of dimer
models \cite{Kasteleyn61}, we get
\begin{equation}
\label{T_d_of_Delta}
\frac{\Delta}{\xi_d T^2}=\frac{2}{T_d^2}-\frac{4G}{\pi}+
\frac{4}{\pi}\int_0^{e^{-T_d^{-2}/2}} \!\!\!\!\! dx \,\, \frac{\arctan(x)}{x},
\end{equation} 
with $\xi_d$ a regularization length for the disorder correlator
\cite{note}, and Catalan's constant $G=0.915966$. With the above
relations and $a=1/\rho=2$ in Eq.~(\ref{eq:K-eff}), we obtain $K(T_d)$
as plotted in Fig.~\ref{fig:1}, together with the data provided by the
dimer algorithm of Ref. \cite{Zeng+99}.  Obviously, there is excellent
agreement between our RBA results and simulations.
\vspace*{-0.3cm}

{\em Finite interaction and persistence length.} --- So far, we have
shown that the model of Eq.~(\ref{H-VS}) with a vanishing single line
persistence length $\xi$ and $U(x)$ representing only a non-crossing
condition is {\em always} in the glassy phase.  We now consider the
effects of an arbitrary repulsive interaction $U(x)$ and a finite
persistence length $\xi$. For the vortex system, $U(x)$ diverges for
$x\to 0$ and decays exponentially for large $x$; $\xi$ can be
identified with the coherence length of the superconductor.

Let us first discuss a general interaction while keeping $\xi=0$.  The
relevance of disorder now depends on the interaction induced change in
$c_{11}$ for $\Delta \to 0$, cf.~Eq.~(\ref{dim}).  This change can be
obtained by employing the known mapping \cite{Pokrovsky+79} of elastic
lines to 1D spin-less fermions which however interact via $U(x)$,
denoting now the finite distance interaction on top of the
self-avoidance.  The compressibility $\kappa$ of the fermions is
related to the compression modulus of the vortex system according to
$1/{\kappa}=a^2 c_{11}$.  For weak interaction, $\kappa$ is easily
calculated, e.g., by bosonization.  Using the standard notation of
Luttinger liquid theory \cite{Voit95}, one gets $\kappa= 1/\pi \hbar
v_N$ with $v_N=v_F+(\tilde U(0)+\tilde U(2\pi/a)/2)/\hbar\pi$, where
$\tilde U(q)$ is the Fourier transform of $U(x)$. Thus, any repulsive
interaction {\em increases} the effective compression modulus
$c_{11}=c_{11}^{\rm PT}+c_{11}^{\rm int}$ with
\begin{equation}
\label{eq:c11-int}
c_{11}^{\rm int}=
\frac{1}{a^2}\left(
\tilde U(0) + \frac{1}{2} \tilde U(2\pi/a)\right),
\end{equation}
and for $U(x) \to 0$, the known Pokrovsky-Talapov result
$c_{11}=c_{11}^{\rm PT}=\pi^2T^2/ga^3$ is recovered. Thus according to
Eq.~(\ref{dim}), the scaling dimension of disorder is {\em increased}
by any finite vortex interaction, and the system remains always in the
glassy phase.

Next, we neglect interactions for the time being and now consider the
thermally fluctuating vortex lines as {\em correlated} random walks,
i.e., we assume a finite persistence length $\xi$.  These correlations
increase the fluctuations of a single line.  Denoting the so decreased
macroscopic line tension again by $g$, entropy reduction due to line
collisions leads not only to the Pokrovsky-Talapov result
but also to an additional renormalization of the
modulus $c_{11}=c_{11}^{\rm PT} - c_{11}^\xi$ where, for $\xi \ll
a^2g/T$,
\begin{equation}
  \label{eq:c11-xi}
  c_{11}^\xi =  \frac{10}{3}\frac{T\xi}{a^2g} c_{11}^{\rm PT} \, .
\end{equation}
For larger $\xi$, $c_{11}$ increases relative to this first order
result. Including the effect of disorder on $c_{11}$,
cf.~Eq.~(\ref{eq:c-11-eff}), one obtains from Eq.~(\ref{dim}) the
{\em finite} transition temperature 
\begin{equation}
  \label{eq:Tg-no-int}
  T_g=0.42\left( \frac{g^2\Delta}{\rho^3\xi} \right)^{1/4}.
\end{equation}
Thus, a finite persistence length yields a finite temperature glass
transition.

Finally, we study the combined effect of interactions and a finite
persistence length. If disorder is strong compared to the interaction,
$\Delta \gg T(\tilde U(0) + \tilde U(2\pi/a)/2)$, the result of
Eq.~(\ref{eq:Tg-no-int}) remains applicable. In the opposite limit,
$T_g$ is independent of disorder,
\begin{equation}
  \label{eq:Tg-int}
  T_g=0.31\left( \frac{g^2(\tilde U(0) + \tilde U(2\pi/a)/2)}
{\rho^3\xi} \right)^{1/3}.
\end{equation}
For vanishing persistence length, $T_g$ diverges, and the system is
glassy at all temperatures.

{\em Discussion.} --- The model of Eq.~(\ref{H-VS}) applies to vortex
systems in films of thickness $d \lesssim a,\lambda$ with $\lambda$
the penetration depth. Then the crucial questions is if the glass
transition temperature $T_g$ can be smaller than the critical
temperature $T_c$ of the superconductor. Since the only mechanism
leading to a finite $T_g$ is a finite persistence length, we maximize
the effect from $c_{11}^\xi$ in Eq.~(\ref{eq:c11-xi}) by considering
small $a \ll \lambda$. Then Ginzburg-Landau theory yields a $\tilde
U(q)$ \cite{Abrikosov64} which is too large for the
bosonization result of Eq.~(\ref{eq:c11-int}) to be valid.  However, a
lower bound is $c_{11}^{\rm int} > b \, c_{11}^{\rm PT}$ with $b$ of
order one.  Using Eq.~(\ref{eq:c11-xi}), we find $T_g > 3b/10 \times
a^2g/\xi$.  With the stiffness $g=\Phi_0^2/16\pi \times d/a\lambda^2$,
$\Phi_0=hc/2e$, \cite{Abrikosov64}, we obtain in the limit
of infinitesimal disorder $T_g \gtrsim \Phi_0^2/16\pi^2 \times
ad/\xi\lambda^2$ which becomes, with $a$, $d > \xi$, $T_g/T_c >
1-5\cdot 10^{-4}$ for $T_c=100$K and with the length $\xi_0=10$\AA,
$\lambda_0=100$\AA\, at $T=0$. Thus, $T_g$ is extremely close to
$T_c$, and presumably any small amount of quenched disorder will raise
$T_g$ above $T_c$, making it impossible to observe a glass transition
in planar vortex systems.

We thank M. Kardar, T. Nattermann, S. Scheidl and C. Zeng for useful
discussions, and especially C. Zeng for sending the data of Fig.~1.
This work was supported by DFG through Emmy-Noether grant No.
EM70/2-1.


\begin{thebibliography}{21}
\bibitem[{not()}]{note}
\bibinfo{note}{The regularization length $\xi_d$ must appear here since a
  discrete and a continuum model are related.}
\end{thebibliography}
\end{document}